\documentclass[12pt]{iopart}
\usepackage{epsfig}  
\begin{document}

\title[Time and length scales in spin glasses]{Time 
and length scales in spin glasses}

\author{L.~Berthier\dag \ddag\ and A.~P.~Young\S}

\address{\dag
Theoretical Physics,
1 Keble Road, Oxford OX1 3NP, UK}

\address{\ddag
Laboratoire des Verres, Universit\'e Montpellier II, 34095
Montpellier, France}

\address{\S
Department of Physics,
University of California,
Santa Cruz, California 95064}

\begin{abstract}
We discuss the slow, nonequilibrium, dynamics of 
spin glasses in their glassy phase.
We briefly review the present theoretical understanding of the 
spectacular phenomena observed in experiments
and describe new numerical results obtained 
in the first large-scale simulation of the nonequilibrium
dynamics of the three dimensional Heisenberg spin glass.
\end{abstract}




\section{Why do we study spin glass dynamics?}
Spin glasses can be seen as one of the paradigms 
for the statistical mechanics of 
impure materials. Experimentally, however, the spin glass
phase is always probed via nonequilibrium dynamic experiments, 
because the
equilibration time of macroscopic samples is infinite. 
Simulations can probe equilibrium behaviour for very moderate sizes
only, so that the thermodynamic nature of the spin glass phase
is still a matter of debate. 
It is also as a model system that the glassy dynamics of spin glasses has 
been studied very extensively in experiments, simulations,
and theoretically in the last two decades~\cite{review1,reviewth}.
Although many theories account for the simplest experimental results,
such as the aging phenomenon, early experiments revealed several other 
spectacular phenomena (rejuvenation, memory, etc.) that are harder
to explain, allowing one to discriminate between various
approaches~\cite{review3}. 

In recent years, several theoretical descriptions 
of the slow dynamics of spin glasses described 
the physics in terms of a distribution
of length scales whose time, $t$, and temperature, $T$, 
evolution depends on the 
specific experimental protocol, as reviewed in Ref.~\cite{review3}.
Aging is described as
the slow growth of a coherence length,
$\ell_T(t)$, reflecting 
quasi-equilibrium/nonequilibrium  at shorter/larger length scales. 
Sensitivity to perturbations of quasi-equilibrated length scales
accounts then for rejuvenation effects, while the strong 
temperature dependence of the growth law $\ell_T(t)$ explains
memory effects~\cite{fh,jp,encorejp,yosh,surf}.
If early numerical studies revealed the existence of such a distribution 
of length scales~\cite{rieger}, its physical relevance 
was critically discussed 
only relatively recently~\cite{yosh2,BB}. 
A major problem, however, is that
most studies focused on the Edwards-Anderson model 
of an Ising spin glass, defined by the Hamiltonian
\begin{equation}
H = - \sum_{<i,j>} J_{ij} S_i S_j,
\label{ea}
\end{equation}
where $S_i = \pm 1$, the sum is over pairs 
of nearest neighbours
of the chosen lattice, and $J_{ij}$ is a quenched random interaction, 
drawn from a symmetric distribution. The interest of
the model (\ref{ea}) is that it has been studied very extensively 
so that some---but only very few!---issues have been settled, 
most notably the existence, for space dimensions $d \ge 3$, 
of a second order phase transition to a spin glass 
phase~\cite{ballesteros,reviewsimu}. 
The nonequilibrium dynamics of the Ising spin 
glass has also been quite extensively studied. Unfortunately, 
for $d=3$, some of the key experimental observations are not 
reproduced~\cite{BB,ricci}, 
although simulations in $d=4$ have been more successful~\cite{BB}. 
This may not be too surprising, since real spin glasses
are made not of Ising spins but vector spins. When the 
interaction between spins is isotropic, the system is 
therefore best described by the Heisenberg spin glass Hamiltonian
\begin{equation}
H = - \sum_{<i,j>} J_{ij} {\bf S}_i \cdot {\bf S}_j,
\label{hsg}
\end{equation}
where the ${\bf S}_i$ are now three-component vectors of unit length. 
The Heisenberg spin glass has been far less studied than 
the Ising one, both statically and dynamically, 
presumably because it was hoped that the understanding 
of the apparently simpler Ising case
would be sufficient to interpret experiments.
Very recent experiments systematically comparing Ising 
(i.e. very anisotropic) and Heisenberg
samples have shown substantial quantitative differences between the two
types of samples, the nonequilibrium effects 
being indeed much clearer in Heisenberg samples~\cite{dupuis,yosh3}.

Hence, it can reasonably be hoped that dynamic studies
of the Heisenberg spin glass in $d=3$ will 
reproduce the key experimental effects, so that
deeper theoretical knowledge of the nature 
of the nonequilibrium dynamics of spin glasses can be gained.
In this paper, we extract some preliminary results from the first
large-scale numerical simulation  of the nonequilibrium dynamics
of the three-dimensional Heisenberg spin glass~\cite{prep}. 

\section{Simulation details}

We simulate the Heisenberg spin glass (\ref{hsg}) in $d=3$.
The sum in (\ref{hsg}) runs over nearest neighbours 
of a cubic lattice with periodic boundary conditions. 
We use a heat-bath algorithm~\cite{peterold} in which 
the updated spin has the correct Boltzmann distribution for the instantaneous
local field. This method has the advantage that a change in the spin
orientation is always made.
We use a rather large 
simulation box of linear size $L=60$, and study
several temperatures $T=0.16$, 0.15, 0.14,
0.12, 0.10, 0.08, 0.04 and 0.02.
Although all the quantities we shall study 
are self-averaging, we use several 
realizations of the disorder, typically 15, to 
increase the statistics of our data. 

Contrary to the Ising case, the
phase transition of the Heisenberg spin glass is still an 
open problem.
A decoupling between spin and chiral degrees of freedom
was theoretically suggested~\cite{kawa}, while early
simulations even questioned the mere existence
of a phase transition~\cite{peterold}.
Very recent simulations involving the most efficient 
tools used to study the Ising spin glass, conclude 
that the model is characterized by a phase 
transition, at $T_c \simeq 0.16$, where both
spins and chirality simultaneously freeze~\cite{peternew}. 
This motivates our choice
for the upper temperature studied in our dynamical 
approach and our use of spin variables as dynamical objects 
of study. 

\section{What has to be measured?}

\begin{figure}
\begin{center}
\psfig{file=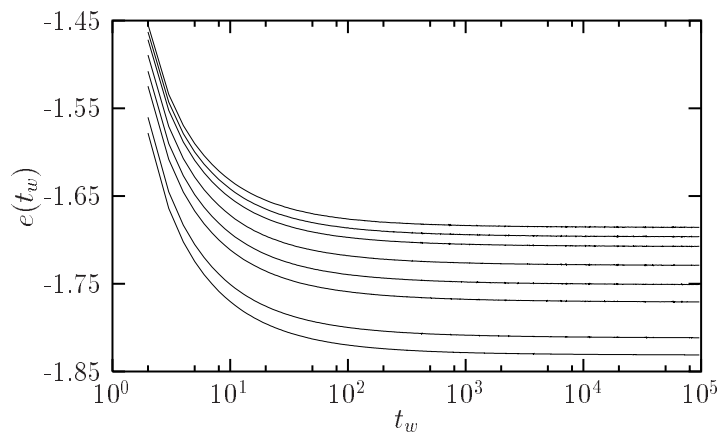,width=9.5cm}
\caption{\label{ener} Time evolution of the energy density (\ref{dens})
after a quench from infinite temperature at the initial
time $t_w$ for $T=0.16$, 0.15, 0.14,
0.12, 0.10, 0.08, 0.04 and 0.02 (from top to bottom).}
\end{center}
\end{figure}

Before embarking in the complex phenomenology of spin glasses, 
it is worth discussing the simplest protocol one can think of
to probe the spin glass phase. A `simple
aging' experiment consists of a sudden quench 
at initial time $t_w=0$ from a
temperature well in the paramagnetic phase, $T \gg T_c$, to a constant, low 
temperature below the spin glass transition, $T < T_c$. 
Aging means a very slow evolution
with time $t_w$ (called `age') of the physical properties 
of the system.
To study this behaviour, we record two types
of quantities. First, `one-time' quantities can be studied, such as 
the energy density of the spin glass,
\begin{equation}
e(t_w) = \frac{1}{N} H.
\label{dens}
\end{equation}
The time evolution of $e(t_w)$ for various low temperatures
is presented in Fig.~\ref{ener}, from which the slow decrease
of the energy towards an asymptotic equilibrium value
is indeed observed, the sign that the dynamics is non-stationary.

\begin{figure}
\begin{center}
\psfig{file=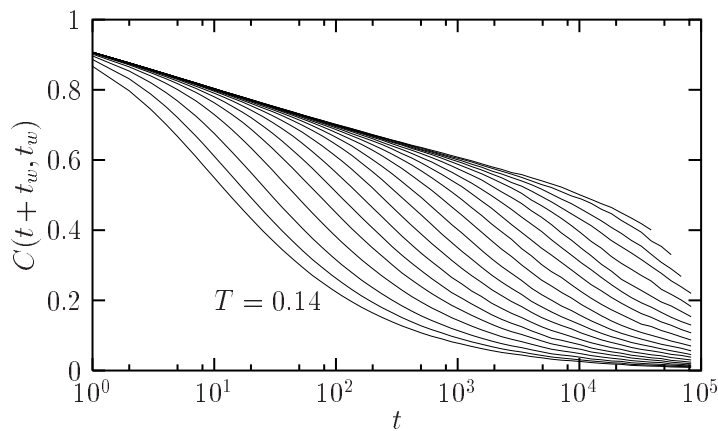,width=9.5cm}
\caption{\label{corr14} Spin-spin autocorrelation
function (\ref{auto}) as a function of the time difference
$t$ for various $t_w$ logarithmically spaced in the 
interval $t_w \in [2,57797]$; $t_w$ increases 
from left to right. The temperature 
is $T=0.14$.}
\end{center}
\end{figure}

We also study `two-time' dynamic quantities. 
While experiments usually record response functions, 
it is easier to measure the corresponding correlation
functions in numerical work. Here, we record the spin-spin
autocorrelation function defined as
\begin{equation}
C(t+t_w,t_w) = \frac{1}{N} \sum_i {\bf S}_i (t+t_w) \cdot {\bf S}_i(t_w).
\label{auto}
\end{equation}
The qualitative behaviour of this function is well-known, and 
a prototypical example is shown in Fig.~\ref{corr14}.
As usual,  
the time decay of $C(t+t_w,t_w)$ can be decomposed 
into two parts. For short time separations, $t \ll t_w$, 
the dynamics is almost independent of $t_w$, 
while the later decay, $t \gg t_w$, becomes slower 
the larger $t_w$. 
Non-stationarity is reflected in the fact 
that $C(t+t_w,t_w) \neq C(t)$.
The physical interpretation
is simple: since the relaxation time 
of the sample is infinite, the only relevant 
time scale is the age of the sample $t_w$ which 
imposes an age-dependent relaxation time: the older the sample, 
the slower its relaxation becomes. 
A careful analysis of short and large time behaviours 
of $C(t+t_w,t_w)$ and comparison to experimental data 
is described in Ref.~\cite{prep}.

\section{Understanding aging in real space}

The key problem is to understand the subtle slow 
changes that the system undergoes: what does `old' or `young' 
really mean for the sample? 
The answer necessarily connects to equilibrium, since
the system eventually equilibrates 
for $t_w \to \infty$. 
Moreover, the decomposition of the decay
$C(t+t_w,t_w)$ between a fast stationary process and a slow
non-stationary one directly suggests the existence of some 
sort of local equilibrium within the sample: a spin appears locally
equilibrated (short-time dynamics) although the sample as a whole
is still far from equilibrium and evolves towards
equilibrium (long-time dynamics). 

\begin{figure}
\begin{center}
\begin{tabular}{ccc}
\psfig{file=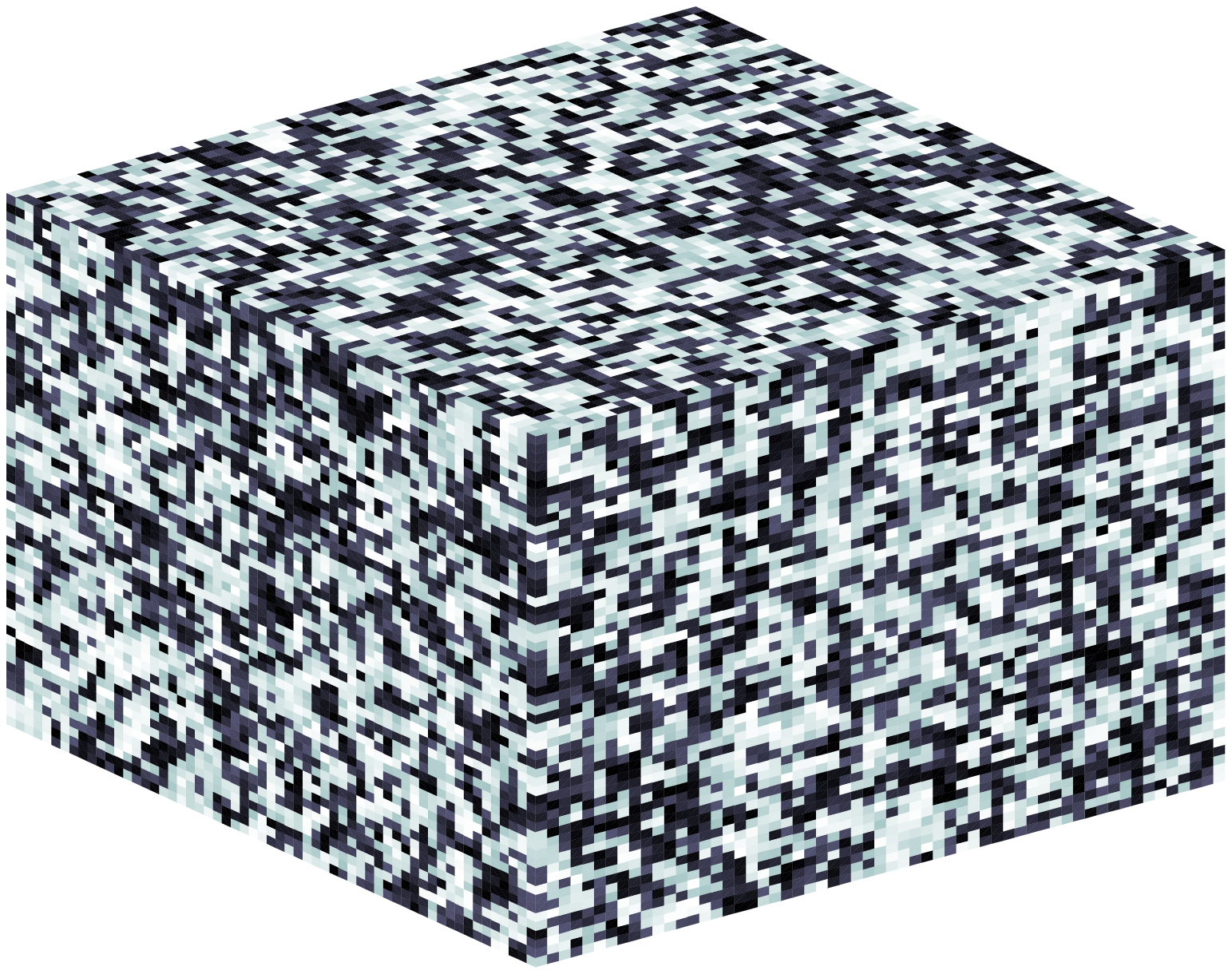,width=4.8cm,height=4.8cm} &
\psfig{file=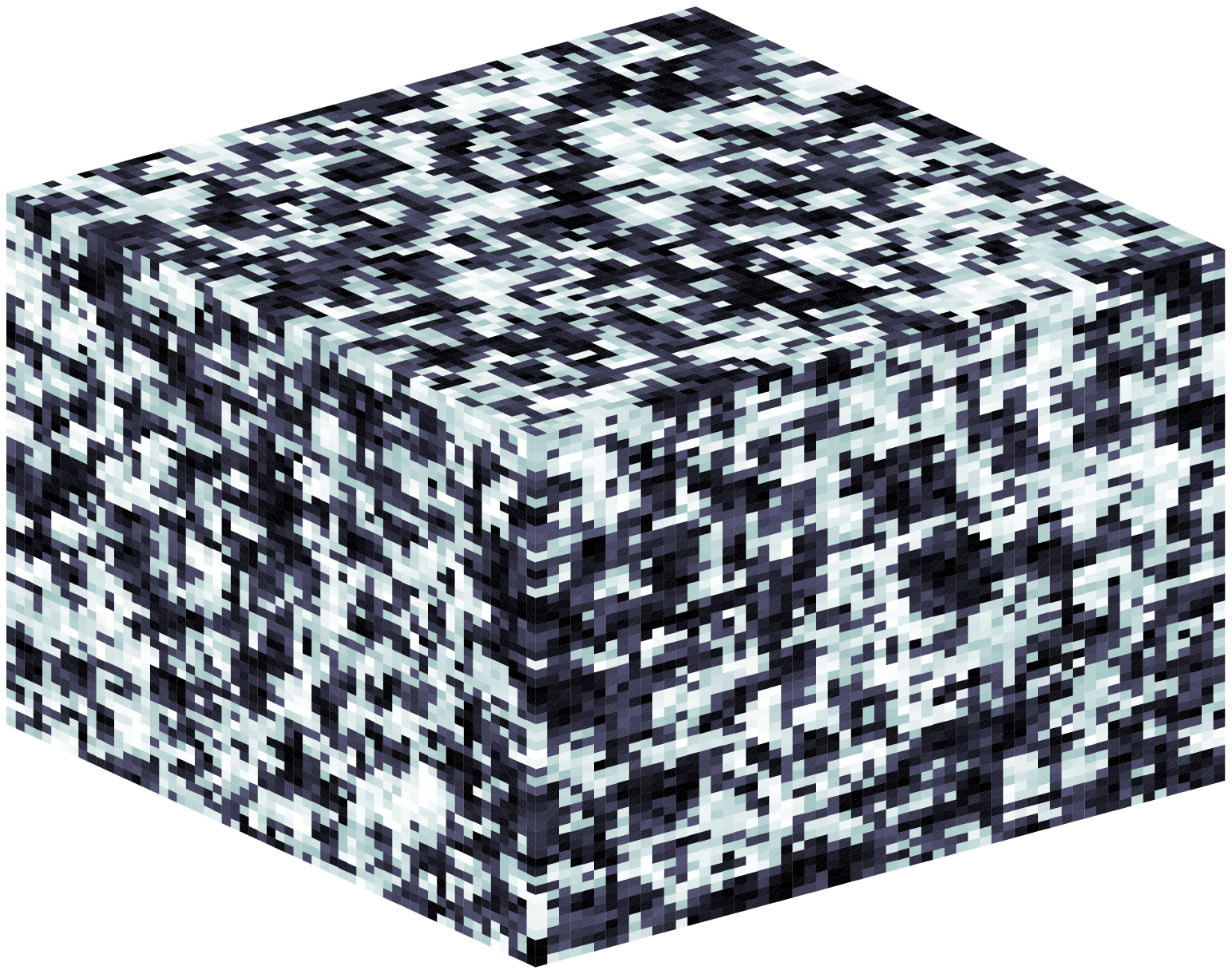,width=4.8cm,height=4.8cm} &
\psfig{file=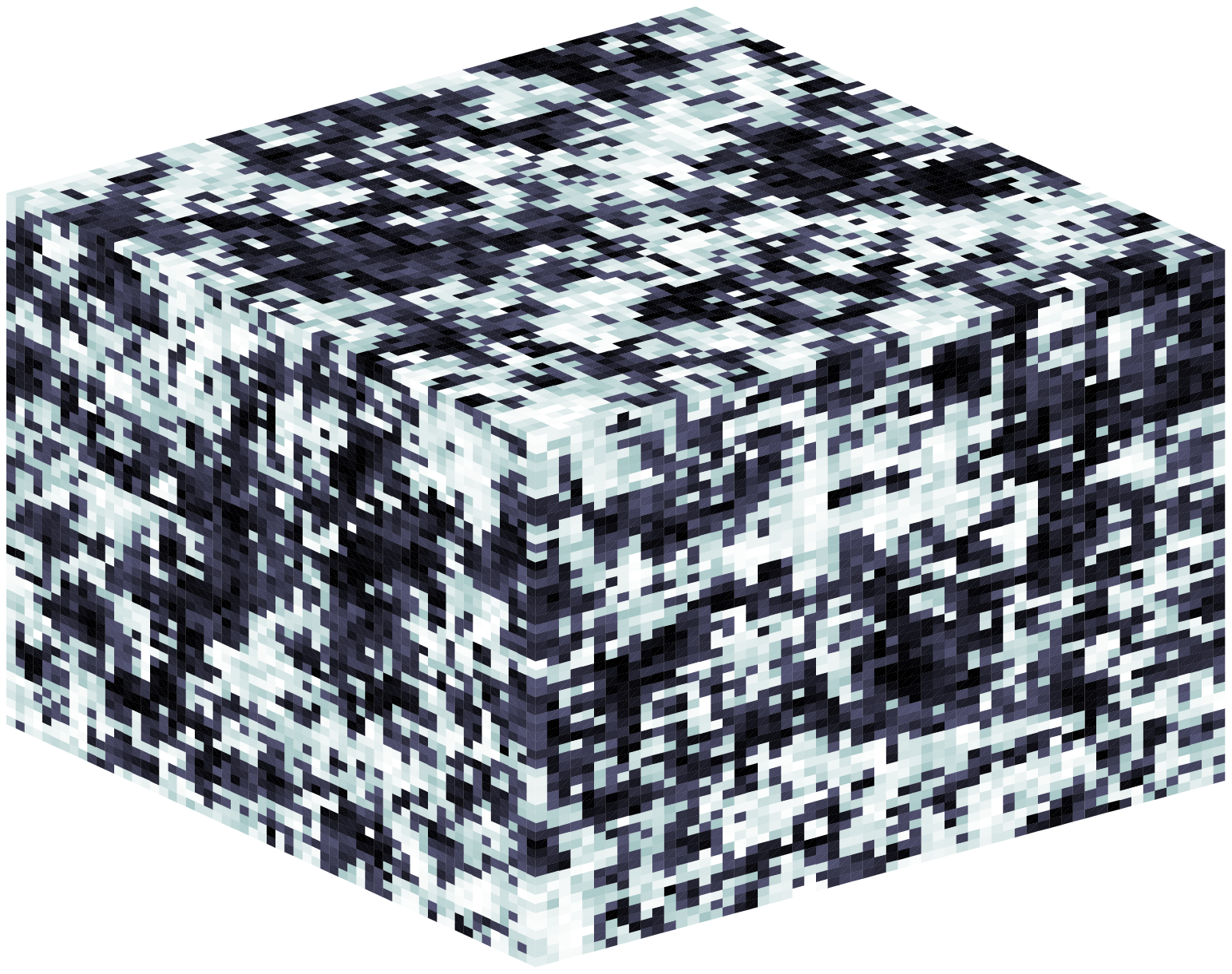,width=4.8cm,height=4.8cm} 
\end{tabular}
\caption{\label{pic}
The orientation variable $\cos \theta_i$
defined in Eq.~(\ref{cos}) is encoded on a grey scale in
a $60 \times 60 \times 60$
simulation box at three different times 
$t_w=2$, 27 and 57797 (from left to right) and 
temperature $T=0.04$. The growth of a local random ordering
is evident.}
\end{center}
\end{figure}

It is possible to illustrate this last statement, as was 
done in the Ising case~\cite{rieger}. Because of the disorder, 
the spin orientations in an equilibrium
configuration are random, so that it is impossible to
detect any domain growth by simply looking at the spin
directions. However, two systems, $(a,b)$, evolving independently but with
the same realization of the disorder will reach correlated
equilibrium configurations~\cite{reviewsimu}, so that the orientation
of the spins in the two copies will be similar, up to a global rotation.
In Fig.~\ref{pic}, we present pictures 
where the `orientation' variable 
\begin{equation}
\cos \theta_i (t) = {\bf S}_i^a (t) \cdot {\bf S}_i^b (t)
\label{cos}
\end{equation}
is encoded on a grey scale. Comparing three successive
times, it becomes clear that aging involves the
growth with time of a local random ordering imposed
by the disorder of the Hamiltonian.  

\begin{figure}
\begin{center}
\psfig{file=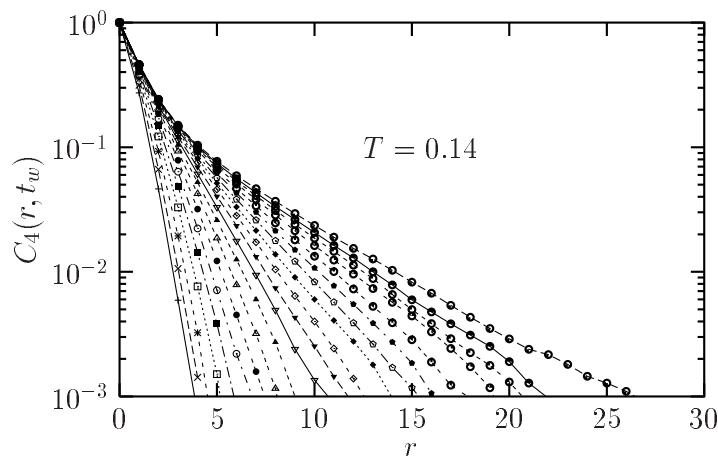,width=9.5cm}
\caption{\label{cav14} Two-spin, two-replica 
correlation function (\ref{c4}) as a function of the 
distance $r$ between the spins 
for various $t_w$ logarithmically spaced in the 
interval $t_w \in [2,57797]$;
$t_w$ increases 
from left to right. The temperature 
is $T=0.14$.}
\end{center}
\end{figure}

It is of course possible to go beyond simple pictures 
of black and white domains and measure 
the growing coherence length $\ell_T(t)$ corresponding to 
the mean domain size in Fig.~\ref{pic}.
For this purpose, the spatial decay of the
following correlation function is recorded: 
\begin{equation}
C_4(r,t) = \frac{1}{N} \sum_i {\bf S}_i^a (t)  \cdot {\bf S}_{i+r}^a (t) \, 
{\bf S}_i^b (t) \cdot {\bf S}_{i+r}^b (t).
\label{c4}
\end{equation}
This function is a straightforward generalization 
of the two-spin, two-replica correlation function 
studied in the Ising case which measures spatial
correlations of the random relative orientation
of two spins~\cite{rieger}. 
In Fig.~\ref{cav14}, we show this function for the 
same parameters as the correlators of Fig.~\ref{corr14}. 
The spatial decay of $C_4(r,t)$ is clearly slower
for larger $t$, in agreement with the pictures in Fig.~\ref{pic}.
Physically, this means that a larger time $t_w$ implies a slower relaxation
due to a larger coherence length, very much as in standard coarsening 
phenomena.

Note that due to periodic boundary conditions, the function
(\ref{c4}) in Fig.~\ref{cav14}
is symmetric about $L/2 = 30$. In Ref.~\cite{kawa}, it was argued that
spin and chirality degrees of freedom undergo different aging dynamics
because they are statically 
decoupled. The numerical support for this statement
was the observation,
for a system of linear size $L=15$, that the autocorrelation
(\ref{auto}) becomes stationary at large $t_w$. 
From Fig.~\ref{c4}, we immediately recognize that the data 
of Ref.~\cite{kawa} are plagued by severe finite size effects, 
so that the conclusions of previous aging studies 
of the Heisenberg spin glass~\cite{kawa,matsu} 
must be taken with some care and
justifies our numerical effort of simulating a very large 
system, $L=60$. The scaling properties of $C_4(r,t)$ and the properties
of the coherence length are further discussed in Ref.~\cite{prep}. 
We make here the important remark that much larger
length scales can be reached in the same numerical time window
for the Heisenberg spin glass than for the Ising case, which may indicate that
richer behaviour can be seen in non-equilibrium simulations
of the Heisenberg spin glass
than the Ising one. 

\section{Conclusion}

We have motivated the need for large-scale numerical simulations 
of the three dimensional Heisenberg spin glass in order
to fill the gap between spatial theoretical descriptions
of spin glass dynamics and experimental observations. 
The results presented here for the dynamics of the model
(\ref{hsg}) show that spin variables qualitatively follow 
the same type of aging behaviour as in the Ising case, which
is due to the slow growth with time of a dynamic coherence length.
In Ref.~\cite{prep}, we analyze in detail the scaling properties of the
dynamic functions reported here. The observation that very large 
length scales can be reached in the numerical time window, see
Fig.~\ref{cav14}, gives us the  
hope, also confirmed by preliminary work, 
that the model will allow us to reproduce most of the 
experimental effects, with the advantage that simulations
have direct access to the distributions of length scales
involved in phenomenological theories, gaining
further understanding of spin glass dynamics.

\end{document}